\documentclass[twocolumn,showpacs,preprintnumbers,amsmath,amssymb]{revtex4}

\usepackage{graphicx}
\usepackage{dcolumn}
\usepackage{bm}

\begin{document}

\preprint{}

\title{ Free flux flow resistivity in strongly overdoped high-$T_c$ cuprate; purely viscous motion of the vortices in semiclassical $d$-wave superconductor}

\author{Y.~Matsuda, A.~Shibata and K.~Izawa}

\affiliation{Institute for Solid State Physics, University of Tokyo, Kashiwanoha 5-1-5, Kashiwa, Chiba 277-8581, Japan}%
\author{H.~Ikuta}
\affiliation{Center for Integrated Research in Science and Engineering, Nagoya University, Furo-cho,Chikusa-ku,Nagoya 464-8603, Japan}%
\author{M.~Hasegawa}
\affiliation{Institute for Materials Research, Tohoku University,  Sendai 980-8577, Japan}
\author{Y.~Kato}
\affiliation{Department of Basic Science, University of Tokyo, Komaba 3-8-1, Meguro-ku, Tokyo 153-8902, Japan}%


\begin{abstract}

	We report the free flux flow (FFF) resistivity associated with a purely viscous motion of the vortices in moderately clean $d$-wave superconductor Bi:2201 in the strongly overdoped regime ($T_c$=16~K) for a wide range of the magnetic field in the vortex state.  The FFF resistivity is obtained by measuring the microwave surface impedance at different microwave frequencies.  It is found that the FFF resistivity is remarkably different from that of conventional $s$-wave superconductors. At low fields $(H<0.2H_{c2})$ the FFF resistivity increases linearly with  $H$ with a coefficient which is far larger than that found in conventional $s$-wave superconductors.  At higher fields, the  FFF resistivity increases in proportion to $\sqrt H$ up to $H_{c2}$.   Based on these results, the energy dissipation mechanism associated with the viscous vortex motion in "semiclassical" $d$-wave superconductors with gap nodes is discussed.  Two possible scenarios are put forth for these field dependence; the enhancement of the quasiparticle relaxation rate and the reduction of the number of the quasiparticles participating the energy dissipation in $d$-wave vortex state.
\end{abstract}

\maketitle

\section {Introduction}
	When a vortex line in a type-II superconductor moves in the superfluid, the frictional force is determined by the damping viscosity, which in turn depends on the energy dissipation processes of quasiparticles.  The problem of the energy dissipation associated with the  viscous motion of the vortices has continued much attention of researchers for years.   To gain an understanding on the energy dissipation, the experimental determination of the free flux flow (FFF) resistivity is particularly important.   Hereafter the term FFF will refer to a purely viscous motion of the vortices, which is realized when the pinning effect on the vortices is negligible.  The FFF resistivity is known to be one of the most fundamental quantities in the superconducting state.   In fully gapped $s$-wave superconductors, the flux flow state has been extensively studied and by now a rather good understanding on the the energy dissipation processes has been achieved. \cite{bs,larkin,kim,bla,stone,esh1,kopnin} In $s$-wave superconductors, the quasiparticles trapped inside the vortex core play a key role in the dissipation processes.  Moreover, it has been shown that there is a fundamental difference in the quasiparticle energy relaxation processes among dirty ($\xi>\ell$), moderately clean ($\xi<\ell<\xi \cdot \frac{\varepsilon_F}{\Delta}$) and superclean ($\ell>\xi \cdot \frac{\varepsilon_F}{\Delta}$) $s$-wave superconductors, where $\xi$ is the coherence length,  $\ell$ is the mean free path, $\varepsilon_F$ is the Fermi energy, and $\Delta$ is the superconducting energy gap.  
  
  A renewed interest in the problem concerning the quasiparticle dissipation owes to recent developments in the investigation of unconventional superconductors.  The latter are characterized by superconducting gap structures which have nodes along certain crystal directions.    In the last two decades unconventional superconductivity has been found in several heavy fermion, organic and oxide materials.   From the viewpoint of the physical properties of the vortex state, perhaps the most relevant effect of the nodes are the existence of gapless quasiparticles extending outside the vortex core.\cite{vol,maki,ichioka1}  In fact recent studies of heat capacity\cite{c}, thermal conductivity\cite{tc}, and NMR relaxation rate\cite{nmr} provide a strong evidence that these quantities are governed by delocalized quasiparticles.  However, despite these extensive studies of the vortex state of unconventional superconductors, the microscopic mechanisms of the energy dissipation associated with the viscous vortex motion is still far from being completely understood,  exposing explicitly our incomplete knowledge of vortex dynamics in type-II superconductors.  Thus it is particularly important to clarify whether the arguments of the energy dissipation are sensitive to the symmetry of the pairing state.\cite{kato} 
  
     Recently, the flux flow resistivities in $f$-wave superconductor UPt$_3$ and $d$-wave high-$T_c$ cuprates, both with line nodes, were demonstrated to be quite unusual.    However,  these materials may not be suitable for the study of the typical behavior of the flux flow resistivity in unconventional superconductors.  The $T$ vs $H$ phase diagram of UPt$_3$, which still is controversial, is considered to consist of various phases with different superconducting gap functions, which complicates considerably the interpretation of the FFF resistivity.\cite{kambe,lut}  The flux flow resistivities of YBa$_2$Cu$_3$O$_{7-\delta}$   and Bi$_2$Sr$_2$CaCu$_2$O$_{8+\delta}$ in the underdoped and optimally doped regimes have been measured by several groups  but here again there are several difficulties in interpreting them.\cite{kun,ym1,tsuchi,ym95,park,hana1}  For instance, the measurements could not cover a wide field range in the vortex state due to extremely large upper critical field $H_{c2}$.   Moreover, very recent STM measurements have demonstrated that the vortex core structure of these high-$T_c$ cuprates is very different from that expected in the semiclassical $d$-wave superconductor\cite{fis,pan},  possibly due to the extremely short coherence lengths and the strong antiferromagnetic fluctuation effect within the core.  
   
   The situation therefore calls for the need for a textbook example of the FFF resistivity of unconventional superconductors with nodes, in which the semiclassical description of the vortex core discussed in the literature, {\it e.g.} Refs.[\onlinecite{vol,maki,ichioka1,kv}] applies.  Especially the FFF resistivity in the "semiclassical" superconductors in the moderately clean regime is strongly desired, because almost all  unconventional superconductors fall within this regime.   It should be noted that the determination of the FFF resistivity is not only important for understanding the electronic structure in the vortex state but is also relevant for analyzing the collective motion of the vortices, such as the flux creep phenomena.  This is easily understood if one recalls that the motion of the vortices in the vortex liquid and solid phase in high-$T_c$ cuprates has been analyzed by assuming the Bardeen-Stephen relation for individual vortex, as discussed in \S V.  Then, if the FFF resistivity strongly deviates from the Bardeen-Stephen relation, the interpretation of the collective motion of the vortices should be modified.   
   
   We stress here that high-$T_c$ cuprates in strongly overdoped regime are particularly suitable for the above purpose because of the following reasons.  (i) Most importantly, it appears that the semiclassical description of the electronic structure of the vortex core is adequate in strongly overdoped materials. \cite{vol,maki,ichioka1,kv}.  This is because many experiments have revealed that in the overdoped regime the electron correlation  and antiferromagnetic fluctuation effects,  which might change the vortex core structure dramatically as observed in STM measurements,   are much weaker than those in optimally doped and underdoped materials.  In fact most of the physical properties in the overdoped materials are well explained within the framework of the Fermi liquid theory.   (ii) Low $H_{c2}$ enables us to measure the FFF resistivity for a wide range of the vortex state.  (iii) The large coherence lengths and small anisotropy ratio reduce the superconducting fluctuation effect which make the interpretation of flux flow resistivity complicated.  In fact, as we discuss in \S IV, the resistive transition of the overdoped materials in magnetic field is much sharper than that of optimally doped and underdoped materials.  (iv) The flux flow Hall angle which complicates the analysis of the flux flow state is very small.\cite{ym1,harr}
   
	The purpose of this work is to present and discuss our experimental results on the FFF resistivity $\rho_f$ of moderately clean $d$-wave superconductors.   The experiments were carried out using strongly overdoped Bi:2201.  This system is an excellent choice for studying the FFF resistivity.  It has a comparatively simple crystal structure (no chain, single CuO$_2$ layer) and hence the band structure is simple.   $H_{c2}$ is within laboratory reach over a very broad range of temperatures.  A major cause of difficulty in obtaining the FFF resistivity in high-$T_c$ cuprates was the strong pinning effect.  To overcome this difficulty, we have measured the microwave surface impedance at different frequencies.    High frequency methods are suitable for this purpose because they probe vortex response at very low currents when the vortices undergo reversible oscillations and they are less sensitive to the flux creep.\cite{gol,cc}   We show that the FFF resistivity of the "semiclassical" $d$-wave superconductor is very different from that of conventional $s$-wave superconductors.  On the basis of the results, we discuss the dissipation mechanism associated with viscous  motion of the vortices in unconventional superconductors.

\section {Experiment}
   
   High quality single crystals of Bi:2201  (Bi$_{1.74}$Pb$_{0.38}$Sr$_{1.88}$Cu$_{1.00}$O$_y$) in the overdoped regime with transition temperature $T_c=16$~K were grown by the floating zone method.\cite{chong}   The sample size used for the microwave measurement was  $\sim$0.8mm$\times$0.7mm$\times$0.04~mm.    The upper inset of Fig.~1 depicts the magnetization at the superconducting transition for the same sample used for the microwave measurements.   The normal state resistivity in the $ab$-plane $\rho_n$  depends on $T$ as $\rho_n\propto T^{\beta}$ with $\beta\sim 2$; the typical Fermi liquid behavior which can be seen in the overdoped high-$T_c$ cuprates.  The resistive transition of the sample in the same batch with $T_c$=18~K is also shown in Fig.~1.    Both resistive transition in zero field and magnetization measurements show a sharp superconducting transition. 

\begin{figure}
\includegraphics [scale=0.5] {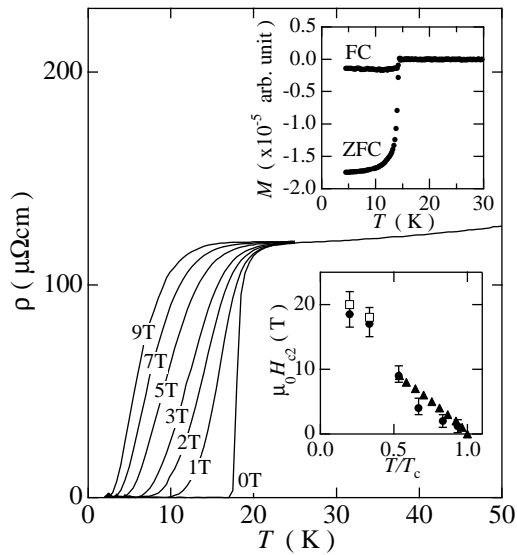}
\caption{The resistive transition in magnetic field of overdoped Bi:2201 in the same batch with $T_{c}$=18~K.  Inset (upper): The magnetization at 5~Oe of the same sample used for the microwave measurements under the conditions of zero field cooling (ZFC) and field cooling (FC).  Inset (lower); $T$-dependence of $H_{c2}$ determined by three different methods.  The filled triangles denote $H_{c2}$ defined by the dc-resistive transition in the main panel, using  criteria $\rho=0.5\rho_n$.    The filled circles denote $H_{c2}$ defined by the magnetic field at which $\rho_1$ becomes frequency independent. The open squares denote $H_{c2}$ defined by the field at which $R_s$ reaches to a normal state value.    $H_{c2}$ is estimated to be $\sim$ 20~T below 5~K. }
\end{figure}

   The microwave surface impedance $Z_s = R_s+i X_s$, where $R_s$ and $X_s$ are the surface resistance and surface reactance, respectively, was measured by the standard cavity perturbation technique using cylindrical cavity resonators made by oxygen free Copper operated in TE$_{011}$ mode.  The resonance frequencies of these cavities were approximately 15~GHz, 30~GHz, and 60~GHz.  The sample was placed in an antinodes of the oscillatory magnetic field $H_{ac}$, such that $H_{ac}$ lies parallel to the $c$-axis of the sample.   The external dc-magnetic field was applied perpendicular to the $ab$-plane.  In this configuration, the two dimensional pancake vortices respond to an oscillatory driving current induced by $H_{ac}$ within the $ab$-planes.  The cavities at 15~GHz and 30~GHz were operated at 1.7~K and sample temperatures were controlled by hot finger techniques using sapphire rod.  The sample temperature in the cavity at 60~GHz was controlled by changing the temperature of the cavity.   The $Q$-values of each cavity are 6.2x10$^4$ for15~GHz, 2.3x10$^4$ for 30~GHz at 4.2~K , and 2x10$^4$ at 4.2~K and 1.5x10$^4$ at 20~K for 60~GHz.  According to the cavity perturbation theory, $R_s$ and $X_s$ can be obtained by 
 \begin{equation}
R_s=G\left(\frac{1}{2Q_s}-\frac{1}{2Q_0}\right)=G\Delta\left(\frac{1}{2Q}\right),
\end{equation}
and
\begin{equation}
X_s=G\left(-\frac{f_s-f_0}{f_0}\right)+C=G\left(-\frac{\Delta f}{f_0}\right)+C,
\end{equation}
where $Q_s$ and $f_s$ are the $Q$-factor and the resonance frequency of the cavity in the presence of the sample, and $Q_0$ and $f_0$ are those without sample.  $G$ is a geometrical factor and $C$ is a metallic shift constant.  

\begin{figure}[t]
\includegraphics [scale=0.5] {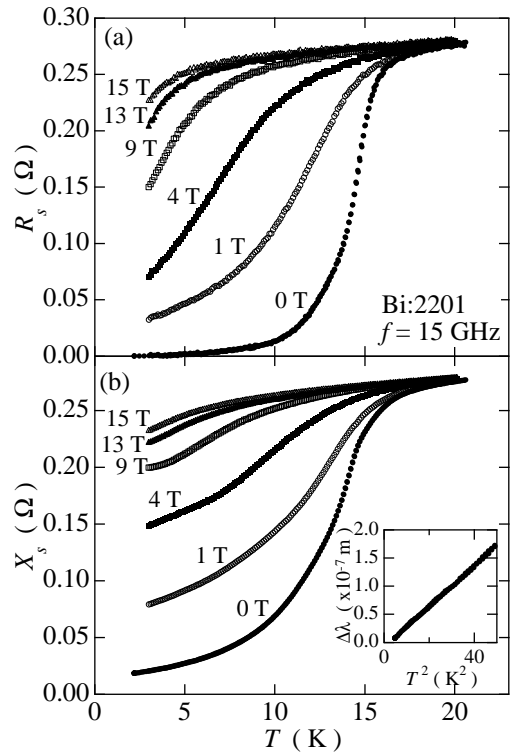}
\caption{$T$-dependence of the surface resistance $R_s$ (a) and surface reactance $X_s$ (b) at 15~GHz in magnetic field.  Both the microwave magnetic field {\boldmath $H$}$_{ac}$ and dc magnetic field {\boldmath $B$} are applied parallel to the $c$-axis ({\boldmath $H$}$_{ac}$$\parallel$ {\boldmath $B$}$\parallel c$).  In this configuration, the energy dissipation is caused by the oscillation of the two-dimensional pancake vortices.   The measurements have been done under the field cooling condition.  The absolute value of $R_s$ and $X_s$ were determined by the normal state dc resistivity.  Inset:  $\Delta \lambda=\lambda (0)-\lambda (T)$ at low temperatures is plotted as a function of $T^2$. }
\end{figure}

	In Figs. 2(a) and (b), the $T$-dependence of $R_s$ and $X_s$ for Bi:2201 at 15~GHz are shown.  The measurements in magnetic field have been performed in the field cooling condition.    We first discuss $R_s$ and $X_s$ in zero field.  In zero field, both $R_s$ and $X_s$ decrease rapidly with decreasing $T$ below the transition.    Let us quickly recall the behavior of $Z_s$ in the superconductors.  In the normal state, the microwave response is dissipative and $R_s=X_s=\mu_0 \omega \delta$, where $\mu_0$ is the vacuum permeability, $\omega/2\pi$ is the microwave frequency, and $\delta_n=\sqrt{2\rho_n/\mu_0\omega}$ is the normal state skin depth.   In Bi:2201,  $\ell$ is estimated to be $\sim$200 $\AA$, which is well shorter than $\delta_n$ at the onset in our frequency range.   We therefore can determine the absolute value of $R_s$ and $X_s$ from the comparison with $\rho_n$ assuming $R_s=X_s$ (Hagen-Rubens relation).   In the Meissner phase, the microwave response is purely reactive and $R_s\simeq 0$ and $X_s=\mu\omega\lambda_{ab}$, where  $\lambda_{ab}$ is the London penetration depth in the $ab$-plane.  Using $\rho_n=130 \mu \Omega $ cm for Bi:2201 at the onset, we obtained  $\lambda_{ab}$=1500 $\AA$ at $T=0$.  This value is slightly smaller than the penetration depth in YBa$_2$Cu$_3$O$_{7-\delta}$ and Bi$_2$Sr$_2$CaCu$_2$O$_{8+\delta}$.  In the inset of Fig.~2(b), $\Delta \lambda=\lambda (0)-\lambda (T)$ at low temperatures is plotted  as a function of $T^2$.  $\Delta \lambda$ is proportional to $T^2$.   The relation $\Delta \lambda \propto T^2$ has been observed in many high-$T_c$ cuprates and  discussed in terms of the superfluid density in $d$-wave superconductors with the impurity state. \cite{lambda}

\section{surface impedance in the vortex state}
	
\begin{figure}[b]
\includegraphics [scale=0.5] {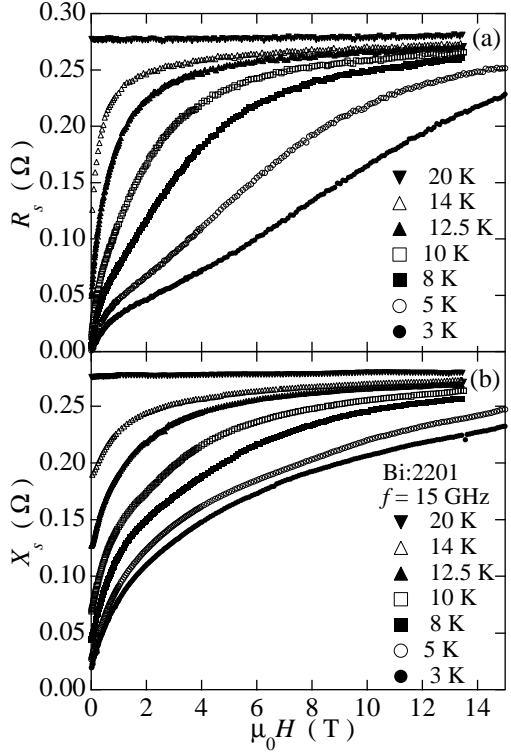}
\caption{Field dependence of the surface resistance $R_s$ (a) and surface reactance $X_s$ (b) at 15~GHz measured by sweeping $H$.  }
\end{figure}

\begin{figure}[b]
\includegraphics [scale=0.5] {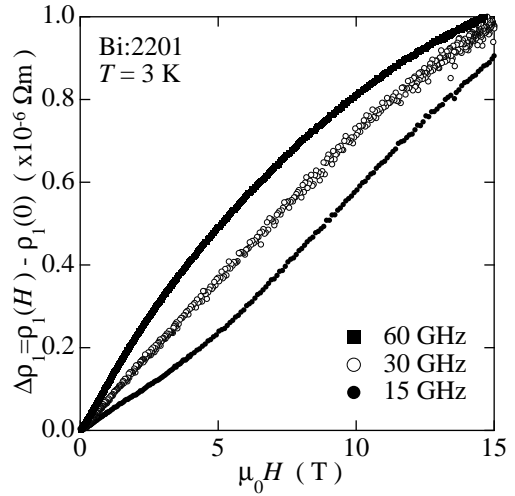}
\caption{The field dependence of $\Delta \rho_1(H)=\rho_1(H)-\rho_1(0)$ obtained from $R_s$ and $X_s$ at three different microwave frequencies.   $\rho_1$ increases with increasing microwave frequency. }
\end{figure}

      We now focus on the surface impedance in the vortex state.  Figure 3 shows the $H$-dependence of $R_s$ and $X_s$ of Bi:2201 at 15~GHz.  In these measurements $R_s$ and $X_s$ are obtained by sweeping $H$.  The hysteresis due to the effect of the trapped field in the crystal is very small.  Moreover, both $R_s$ and $X_s$ obtained by sweeping $H$ well coincide with those obtained under the field cooling conditions shown in Fig.~2.   These results indicate that neither inhomogeneous field distribution inside the crystal nor magnetostriction\cite{ikuta} caused by sweeping $H$ seriously influences the analysis of $Z_s$.  
	
	In the vortex state, $Z_s$ is governed by the vortex dynamics.  We may roughly estimate $R_s$ in the  limit of large and negligible rf field penetration as follows.  In the flux flow state when the pinning frequency $\omega_p/2\pi$ is negligible compared to the microwave frequency ($\omega_p\ll\omega$), two characteristic length scales, namely $\lambda_{ab}$ and the flux flow skin depth $\delta_f\sim \sqrt{2\rho_f/\mu_0\omega}$, appear in accordance with the microwave field penetration.  At low fields, $\lambda_{ab}$ greatly exceeds $\delta_f$ ($\lambda_{ab} \gg \delta_f$).  In this regime, $R_s$ and $X_s$ are given as $R_s\sim\rho_f/\lambda_{ab}$ and $X_s\sim \mu_0\omega\lambda_{ab}$.  On the other hand, at high fields where $\delta_f$ greatly exceeds $\lambda_{ab}$ ($\delta_f \gg \lambda_{ab}$), the viscous loss becomes dominant and the response is similar to the normal state ($R_s\simeq X_s$) except that $\delta_n$ is replaced by $\delta_f$.  In the presence of pinning centers of the vortices, $R_s$ is reduced as discussed below. 
	
   We here analyze the field dependence of $Z_s$ in accordance with the theory of Coffey and Clem.\cite{cc}  The equation of vortex motion for the vortex line velocity {\boldmath$u$},
\begin{equation}
\eta \mbox{\boldmath $u$}+\kappa_p \mbox{\boldmath $x$}=\Phi_0 J \times \hat{\mbox{\boldmath $z$}}
\end{equation}
where $\eta$ and $\kappa_p$ are the viscous drag constant and pinning parameter, respectively, and $\hat{\mbox{\boldmath $z$}}$ the unit vector parallel to {\boldmath $B$} (we take {\boldmath $J$}$\parallel${\boldmath $x$}).  According to Coffey and Clem, the field dependence of $Z_s$ in the Meissner and vortex phases is expressed as
\begin{equation}
	Z_s=i\mu_{0}\omega\lambda_{ab}\left[\frac{1-(i/2){\delta_v}^{2}/ \lambda_{ab}^{2}}{1+2i\lambda_{ab}^{2}/\delta_{nf}^2}\right]^{1/2},
\end{equation} 
where $\delta_v^2=\delta_f^2(1-i\omega_p/\omega)^{-1}$ with $\omega_p/2\pi=\kappa_p/2\pi \eta$ being the pinning frequency.  Writing $Z_s$ in terms of the complex resistivity $\rho=\rho_1+i\rho_2$ as $Z_s=\sqrt{i\omega\mu_0(\rho_1+i\rho_2)}$, we have
\begin{equation}
\rho_1=\mu_0\omega\frac{\lambda_{ab}^2s}{1+s^2}+\rho_f\frac{1}{1+s^2}\frac{1+sp}{1+p^2},
\label{eqn:rho1}
\end{equation}
and
\begin{equation}
\rho_2=\mu_0\omega\frac{\lambda_{ab}^2}{1+s^2}+\rho_f\frac{1}{1+s^2}\frac{p-s}{1+p^2},
\label{eqn:rho2}
\end{equation}
where $s=2\lambda_{ab}^2/\delta_{nf}^2$ and $p=\omega_p/\omega$.   In Eqs.(\ref{eqn:rho1}) and (\ref{eqn:rho2}), the first terms in the right hand side are $\rho_1$ and $\rho_2$ at zero field, and second terms represent the field dependence.   In what follows we discuss the microwave response focusing on $\rho_1$ obtained from $R_s$ and $X_s$.   Figure 4 shows the field dependence of $\Delta \rho_1(H)=\rho_1(H)-\rho_1(0)$ at three different microwave frequencies.    The field dependence of $\Delta \rho_1$ is frequency dependent; $\rho_1$ increases with increasing frequency.  Since $\rho_1$ is reduced by the vortex pinning effect, as seen in Eq.(\ref{eqn:rho1}),  this result indicates that the pinning effect of the vortices is not negligible for the analysis of the flux flow resistivity in our microwave frequency range.  Therefore,  it is necessary to determine the pinning frequency for an accurate determination of the FFF resistivity.  

\begin{figure}[b]
\includegraphics [scale=0.5] {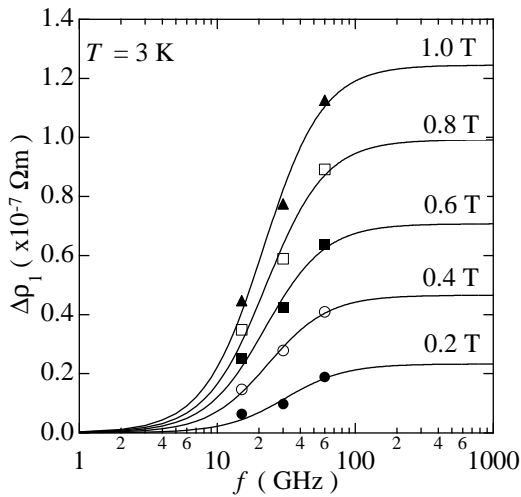}
\caption{Frequency dependence of $\Delta \rho_1(H)$ at $T=$3.0~K.  (Filled triangles (1.0~T), open squares (0.8~T), filled squares (0.6~T), open circles (0.4~T),  filled circles (0.2~T)). The solid lines are the results of the fitting by $\Delta \rho_1(B,\omega)=\rho_f \omega^2/(\omega^2+\omega_p^2)$.   For details, see the text.  }
\end{figure}

	In Fig.~5, $\Delta \rho_1$ at $T$=3~K is plotted as a function of the microwave frequency.   The solid lines show the results of the fitting by $\Delta \rho_1(H,\omega)=\rho_f \omega^2/(\omega^2+\omega_p^2)$.  It should be noted that since $s \ll 1$ except the vicinity of $H_{c2}$, the $H$-dependence of $s$ little influences the present analysis.  Nevertheless, we restrict our analysis at $H\alt$10~T to avoid the influence of $H$-dependence of $s$.   The fitting parameters are $\omega_p$ and $\rho_f$.  The ambiguity for determining $\omega_p$ and $\rho_f$ is small.  The $H$-dependence of the pinning frequency  obtained by the fitting is depicted in Fig.~6.   At low field, $\omega_p/2\pi$ is approximately 22~GHz at $T=$3~K and 17~GHz at 5~K.  These values are much larger than the pinning frequency in Bi$_2$Sr$_2$CaCu$_2$O$_{8+\delta}$ but much smaller than $\omega_p/2\pi$ in YBa$_2$Cu$_3$O$_7$ \cite{tsuchi,hana1,gol}.  At low field, $\omega_p$ decreases gradually, while at $\agt$1.5~T  $\omega_p$ decreases approximately as $\omega_p \propto H^{-1}$, as shown in the inset of Fig.~6.

\section{free flux flow resistivity of Bi:2201}

\begin{figure}[b]
\includegraphics [scale=0.5] {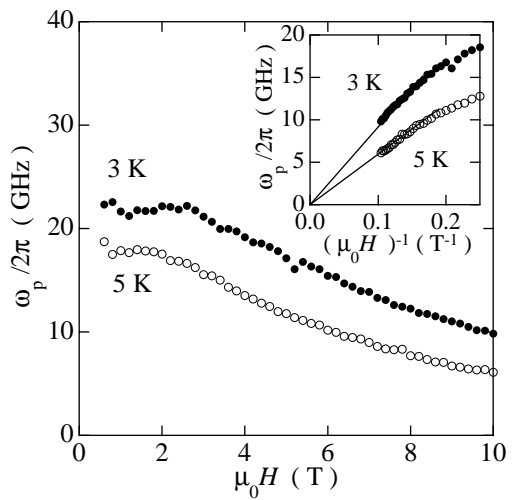}
\caption{The field dependence of the pinning frequency $\omega_p/2 \pi$ at $T=$ 3~K and 5~K obtained by the fitting shown in Fig.~4.  Inset:  Same data plotted as a function of $1/H$.  $\omega_p/2 \pi$ decays in proportion to $1/H$.  The sloid lines show the relation $\omega_p/2 \pi\propto1/H$. }
\end{figure}

	 Before discussing the FFF resistivity,  it will prove useful to first comment on $H_{c2}$ of Bi:2201.   It is well known that the resistive transitions of high-$T_c$ cuprates are significantly broadened in magnetic field due to the strong thermal fluctuation effect and the vortex dynamics.   Although in overdoped Bi:2201 such a broadening effect is relatively small, it still becomes an obstacle in determining $H_{c2}$ \cite{fluc}.   In the lower inset of Fig.~1, we plot $H_{c2}$ determined by three different methods.  The filled triangles represent $H_{c2}$ defined by the dc-resistive transition in Fig.~1, using a criteria $\rho=\frac{1}{2}\rho_n$.  The filled circles are $H_{c2}$ defined by the magnetic field at which $\rho_1$ becomes frequency independent.   The open squares represent $H_{c2}$ defined by the field at which $R_s$ reaches to a normal state value.   The values of $H_{c2}$ obtained from the three different methods do not differ significantly.  A striking divergence in $H_{c2}$ as the temperature approached zero was reported in the overdoped  Tl:2201 in the transport measurements,\cite{mack}  while such a divergent behavior was not observed in the specific heat and Raman scattering measurements.\cite{raman}    The divergent behavior of $H_{c2}$ was discussed in terms of several proposed models, such as the Josephson coupled small grains with $T_c$ higher than the bulk.\cite{gesh,bi2201}   However in the present Bi:2201 such anomalies are not observed in $H_{c2}$ at least above 2~K.  At present we do not know the reason for this difference.  From these measurements,  $H_{c2}$ is estimated to be approximately 20~T below 5~K.    
	
	 In Fig.~7(a), we plot $\rho_f/\rho_n$ as a function of $H/H_{c2}$ at 3~K, assuming $H_{c2}$=19~T.  If we assume $H_{c2}$=17~T at $T=$5~K, both $\rho_f$ almost exactly coincide with $\rho_f$ at 3~K, as shown in Figs.~7(a) and (b).  The field dependence of $\rho_f$ is convex.   We found that the there are two characteristic regimes in the $H$-dependence of $\rho_f$.   In the low field region ($H/H_{c2}<$0.2),  $\rho_f$ increases linearly with $H$ as
\begin{equation}
\rho_f=\alpha\frac{H}{H_{c2}}\rho_n
 \label{eq:dlow}
\end{equation}
with $\alpha \simeq 2 $.    A deviation from $H$-linear dependence is clearly observed at higher field.  In Fig.~7(b), $\rho_f/\rho_n$ is plotted as a function of $\sqrt {H/H_{c2}}$.   We found that $\rho_f$ increases  as 
\begin{equation}
\rho_f \propto \sqrt{\frac{H}{H_{c2}}}
 \label{eq:dhigh}
\end{equation}
at $H/H_{c2}\agt$0.2.  Since the linear extrapolation of $\rho_f/\rho_n$  in Fig~7(b) points to $\rho_f/\rho_n=1$ at $H/H_{c2}$=1,  it is natural to expect that the relation of Eq.(\ref{eq:dhigh}) continues all the way up to $H_{c2}$.  

\begin{figure}[t]
\includegraphics [scale=0.5] {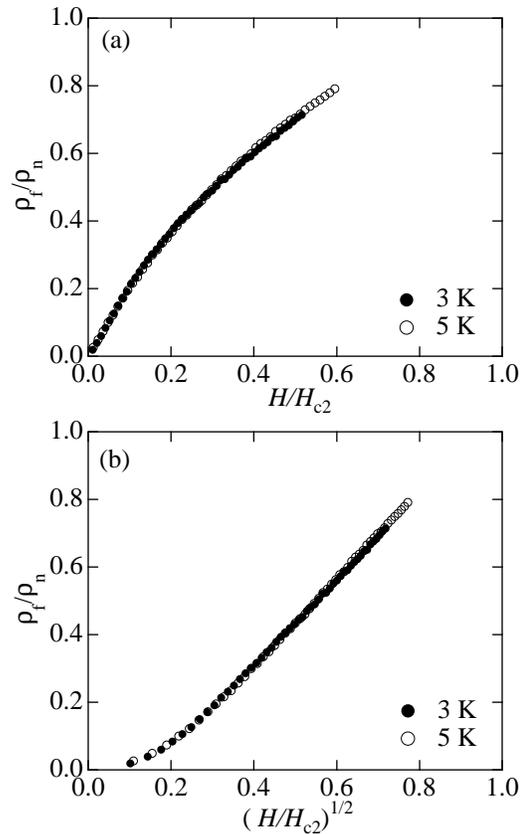}
\caption{(a) The flux flow resistivity at $T=$3~K and 5~K as a function of $H/H_{c2}$.  We assumed $H_{c2}$=19~T at 3~K and 17~T at 5~K.  The flux flow resistivity is normalized by the normal state value.  (b) Same data plotted as a function of $\sqrt{H/H_{c2}}$.   }
\end{figure}

\section{Discussion}
 	
\subsection{Flux flow in $s$-wave superconductors}

	In order to contrast the present results with the FFF resistivity of isotropic $s$-wave superconductors,  we first briefly review the flux flow state in $s$-wave superconductors. 	 
	
	For isotropic $s$-wave pairing in the dirty regime, the Bardeen-Stephen model appears to be quite successful in describing the energy dissipation.\cite{bs,larkin,kim}    The Bardeen-Stephen theory models the vortex core as a cylinder whose radius is the coherence length.  It is assumed that the core is a normal metallic state inside of which  the energy dissipation is dominated by the impurity scattering, similar to the ordinary resistive process.  This is a good approximation for dirty superconductors with $\ell<\xi$.  It follows from this model that the FFF resistivity in dirty $s$-wave superconductors is proportional to the normal state resistivity and is to the number of the vortices, 
 \begin{equation}
 \rho_f=\rho_nH/H_{c2}. 
 \label{eq:sdirty}
 \end{equation} 
The validity of this Bardeen-Stephen relation has been confirmed in most of dirty $s$-wave superconductors almost throughout the whole Abrikosov phase;  $H_{c1}<H<H_{c2}$. \cite{kim} 
 
 	However, the description of vortex core as a normal metal is limited to dirty $s$-wave superconductors.  In moderately clean and superclean $s$-wave superconductors with $\ell>\xi$,  the quasiparticle response to an electromagnetic field is radically different from that of normal electrons, since the model of a normal metallic core breaks down. \cite{esh1,kopnin}  The difference lies in the fact that the quasiparticles in the core are subject to Andreev reflections by the pair potential and form the bound states of Caroli, de Gennes and Matricon \cite{cdm,gygi} before getting scattered by impurities.  The largest energy difference between the bound states is roughly estimated as $\hbar \Omega_0\sim \Delta^2/\varepsilon_F$, where $\Omega_0$ is the angular velocity.  The electric conduction in the vortex state is governed by the scattering time between the Andreev bound states in the presence of impurities.    Effects of these quasiparticles on the vortex dynamics have been considered in a number of papers.   For moderately clean $s$-wave superconductors, the FFF resistivity has been calculated as\cite{larkin,kopnin}
\begin{equation}
\rho_f\sim \rho_n \frac{1}{\ln\left(\frac{\Delta}{k_{B}T}\right)}\frac{H}{H_{c2}}.
 \label{eq:sclean}
\end{equation}
The logarithmic factor results from the shrinkage of the vortex core at low temperature and logarithmic energy dependence of the impurity scattering rate of the Andreef bound state (Kramer-Pesch effect) \cite{kp}.  Thus, in spite of the fundamental difference of the character of the quasiparticles within the vortex core, the FFF resistivity in the moderately clean $s$-wave superconductors increases in proportion to $H$, which is similar to that in the dirty superconductors.   In fact, the FFF resistivity of several  moderately clean $s$-wave superconductors discovered recently were found to be proportional to $H$, though the logarithmic correction at very low temperature has never been reported so far. \cite{izawa}

\subsection{Flux flow in  $d$-wave superconductor}
  
  We are now in position to discuss the FFF resistivity of semiclassical $d$-wave superconductors.  It is obvious from Figs.~7(a) and (b) that {\it the field dependence of $\rho_f$ expressed as Eqs.(\ref{eq:dlow}) and (\ref{eq:dhigh}) is markedly different from that of conventional $s$-wave superconductors expressed as Eqs. (\ref{eq:sdirty}) and (\ref{eq:sclean})}.   	
  
 	We first discuss the low field behavior of the FFF resistivity in Bi:2201.   The linear dependence of $\rho_{f}$ on the magnetic field means that the energy dissipation per vortex does not depend on the magnetic field or the inter-vortex spacing.  We can interpret this fact naturally if the energy dissipation is assumed to occur mainly near each vortex even in the superconductors with gap nodes. In fact, this assumption is justified by a numerical result on the ac response of the $d$-wave vortex.\cite{esh1,esh2}   Comparing  Eq.(\ref{eq:dlow}) with Eq.(\ref{eq:sclean}),  the coefficient of the $H$-linear term in $d$-wave superconductor is found to be nearly as twice as that in $s$-wave superconductors.   This behavior is similar to UPt$_3$ with line nodes, in which $\rho_f$  at low field is larger than that found in conventional $s$-wave superconductors.\cite{kambe}  It should be noted that a similar result was reported in very recent measurements of high purity borocarbide superconductor YNi$_2$B$_2$C with very anisotropic superconducting gap,presumably anisotropic $s$-wave symmetry.\cite{izawa,hana2}    These results lead us to conclude that a large initial slope is a common feature in the FFF resistivity of the superconductors with nodes.   In what follows, we discuss possible origins for the enhancement of the FFF resistivity at low fields on the basis of the theoretical results available at the present stage. 

	According to Kopnin and Volovik, the vortex transport in semiclassical $d$-wave superconductors is governed by dynamics of quasiparticles which form Andreev bound states around a vortex, much like in $s$-wave superconductors.\cite{kv2}  The excitation spectrum of those quasiparticles is given by
\begin{equation}
E(L,\theta)=-\Omega(\theta)L
\end{equation}
in terms of the angle $\theta$ in the momentum space and $L$ the angular momentum. In this expression, $\Omega(\theta)$ denotes the angular velocity, which depends on the direction $\theta$.  Roughly speaking, $\Omega(\theta)$ is proportional to the square of the energy gap, $\Delta(\theta)$ ($\propto \cos(2\theta)$ for $d_{x^2-y^2}$ states). This branch corresponds to the Caroli-de~Gennes-Matricon mode in the isotropic $s$-wave superconductors (in $s$-wave symmetry, $\Omega_0$ is $\theta$ independent).\cite{cdm,gygi}   The quasiparticles with $\theta$ away from the nodes in $d$-wave vortex are well localized near vortex cores and they are similar, in nature, with those in an $s$-wave vortex.  As the angle $\theta$ approaches a nodal direction, however, the quasiparticles become more extended and farther away from the vortex cores.  In this way the character of quasiparticles in $d$-wave vortex is very different from that of quasiparticles in the $s$-wave vortex. 

	According to the theory by Kopnin and Volovik based on the relaxation time approximation, FFF resistivity is given by 
\begin{equation}
\rho_f=\frac{B}{\langle\Omega(\theta)\rangle \tau_{\rm v} n_{\rm e}|e|c},
\label{eq:rhof}
\end{equation}
where $\langle \cdots  \rangle$ denotes the average over the Fermi surface, $\tau_{\rm v}$ is the relaxation time of quasiparticles, and $n_e$ is the carrier density in the vortex state.\cite{kv}   In the theory of the relaxation time approximation,  the transport coefficients are given in the form of the parallel  circuit;  the conductivity is expressed as a sum of the contribution from each part of the Fermi surface.  Then magnitude of resistivity in vortex state expressed by Eq.(\ref{eq:rhof}) is governed by the largest value of $\Omega(\theta)$ on the Fermi surface. This fact is physically interpreted in the following way.  The quasiparticles with smaller $\Omega(\theta)$ come from the vicinity of nodes. They are only weakly excited by vortex motion,  because such quasiparticles are extended in regions far away from vortex cores. On the other hand, quasiparticles with larger $\Omega(\theta)$ are localized near vortex cores. Therefore it is likely that such quasiparticles are excited substantially by vortex motion and an appreciable deviation of the distribution function from the equilibrium state may occur.    Thus when the gap has nodes, portions of the Fermi surface near the nodal directions do not contribute to $\langle\Omega(\theta)\rangle$.  This is in marked contrast to the isotropic $s$-wave superconductors, in which every part of the Fermi surface can contribute to $\langle\Omega(\theta)\rangle$.  The reduction of the number of quasiparticles available for the energy dissipation in the superconductors with nodes gives rise to the enhanced flux flow resistivity. This scenario has been adopted in Ref.[\onlinecite{kambe}] to discuss the flux flow resistivity of UPt$_3$.  Although this argument explains the low field ($H<0.2H_{c2}$) behavior expressed as (\ref{eq:dlow}), it gives no account for the $\sqrt{H}$-dependence of $\rho_f$ expressed as Eq.({\ref{eq:dlow}) observed in the almost whole regime at higher field ($0.2H_{c2}\alt H <H_{c2}$). 

	There is, however, another scenario. In the following part, we show that the reduction of $\tau_{\rm v}$ in $d$-wave vortex states explains consistently both (\ref{eq:dlow}) and (\ref{eq:dhigh}).   Here we regard the impurity scattering as the main process of relaxation in the cuprates.  Within the Born approximation, $\tau_{\rm v}$ is inversely proportional to the density of states (DOS) of quasiparticles available as the out-going states in the scattering process of localized quasiparticles. On the other hand, the low energy DOS of quasiparticles in $d$-wave vortex states are known to be larger than that in $s$-wave vortex states theoretically.\cite{vol,kv,ichioka2}  In Ref.[\onlinecite{kv}], Kopnin and Volovik calculated the density of states $N_{\rm v}(E)$ per each $d$-wave vortex for energy $E$ to obtain 
\begin{equation}
N_{\rm v}(E)\sim N_0 \xi^2 \left(\Delta/E\right)\sim  N_0 \xi r(E), 
\label{eq:dos}
\end{equation}
where $N_0$ denotes the DOS on the Fermi surface in the normal states and $r(E)=\hbar v_F/E$ with $v_F$ the Fermi velocity. The singularity at $E=0$ is removed by a cut-off length. According to Ref.[\onlinecite{kv}], for energy $E$ satisfying $r(E)>R_{\rm B}$ with the intervortex distance $R_{\rm B}\sim \xi\sqrt{H_{c2}/B}$, $r(E)$ should be replaced by $R_{\rm B}$ for {\it pure} superconductors without impurity scattering. For {\it impure and clean} superconductors, instead, we speculate that $r(E)$ should be replaced by $R_{\rm B}$ or the mean free path $l_{\rm v}(=v_F\tau_{\rm v})$, whichever is smaller. We then expect that  
\begin{equation}
N_{\rm v}(0)/(N_0 \xi^2)\sim \left\{
\begin{array}{cc}
l_{\rm v}/\xi,& l_{\rm v}<R_{\rm B}\\
\sqrt{H_{c2}/B},& R_{\rm B}<l_{\rm v}.
\end{array}
\right.
\label{eq:dos2}
\end{equation}
for $E=0$.  The quasiparticle DOS per each isotropic $s$-wave vortex is given by $N_0 \xi_0^2$. \cite{cdm} Therefore, the left-hand side in (\ref{eq:dos2}) gives the ratio of DOS in $d$-wave vortex states to that in the isotropic $s$-wave vortex state. From this fact and Eq. (\ref{eq:dos2}), we expect that
\begin{equation}
\tau_v(d\mbox{-wave})/\tau_v(s\mbox{-wave})
\sim \left\{
\begin{array}{cc}
\xi/l_{\rm v},& l_{\rm v}<R_{\rm B}\\
\sqrt{B/H_{c2}},& R_{\rm B}<l_{\rm v}.
\end{array}
\right
. 
\label{eq:tauratio}
\end{equation}
This reduction of the relaxation time in $d$-wave vortex also yields the enhancement of the flux flow resistivity $\rho_f$. If we assume here that this reduction of $\tau_{\rm v}$ {\it alone} leads to the enhancement of $\rho_f$, i. e. 
\begin{equation}
\rho_f(d\mbox{-wave})/\rho_f(s\mbox{-wave})\sim \tau_v(s\mbox{-wave})/\tau_v(d\mbox{-wave})
\end{equation}
and $\rho_f(s\mbox{-wave})\sim \rho_n \left(B/H_{c2}\right)$, we obtain 
\begin{equation}
\rho_{\rm f}(d\mbox{-wave})/\rho_{\rm n}\sim \left\{
\begin{array}{cc}
\left(l_{\rm v}/\xi\right)\left(B/H_{c2}\right),& l_{\rm v}<R_{\rm B}\\
\sqrt{B/H_{c2}},& R_{\rm B}<l_{\rm v}.
\end{array}
\right. 
\label{eq:rhofdwave}
\end{equation}
We then see the upshot of the hyposesis (\ref{eq:rhofdwave}).  The expression (\ref{eq:rhofdwave}) is consistent with the experimental results on $\rho_{\rm f}$ both in low fields Eq.~(\ref{eq:dlow}) and in high field Eq.~(\ref{eq:dhigh}).  From the relation $l_{\rm v}\sim R_{\rm B}$ at the crossover field 2T$\sim$3T from Eq.~(\ref{eq:dlow}) to Eq.~(\ref{eq:dhigh}), we obtain $l_{\rm v}=280\sim 340\AA$.  From this value of $l_{\rm v}$ and $\xi\sim 42\AA$ (estimated from $H_{c2}=20$T), we obtain $l_{\rm v}/\xi= 6.6\sim 8$. This value is somewhat larger than $\alpha\sim 2$.  With consideration of the crudeness of our estimation, however, we should say that these two values are of the same order.

	At the present state of the study, we do not know whether the dominant source for quasiparticle energy dissipation comes from the reduction of the number of the quasiparticles or the enhancement of the carrier scattering rate.  A detailed numerical calculation for the energy dissipation especially when each vortex overlaps with its neighborhood would be necessary.

\section{summary}	

The microwave surface impedance measurements in the vortex state of overdoped Bi:2201 demonstrate that the free flux flow resistivity in the moderately clean $d$-wave superconductor with gap nodes is remarkably different from that in conventional fully gapped $s$-wave superconductors.  At low fields, the free flux flow resistivity increase linearly with $H$ with a coefficient which is far larger than that found in conventional $s$-wave superconductors.  At higher fields, the flux flow resistivity increases in proportion to $\sqrt H$ up to $H_{c2}$.  Two possible scenarios are put forth for these field dependence; the enhancement of the quasiparticle relaxation rate and the reduction of the number of the quasiparticles participating the energy dissipation in $d$-wave vortex state.  The present results indicates that the physical mechanism of the energy dissipation associated with the purely viscous motion of the vortices are sensitive to the symmetry of the pairing state.   

\section{acknowledgement}	

We thank T.~Hanaguri, T.~Kita, A.~Maeda and A.~Tanaka, for helpful discussions.  This work was supported by a Grant-in-Aid for Scientific Research on Priority Area "Novel Quantum Phenomena in Transition Metal Oxides." from the Ministry of Education, Culture, Sports, Science, and  Technology of Japan.  	
\end{document}